\documentclass[preprint,showkeys,preprintnumbers,amsmath,amssymb]{revtex4}

\usepackage{graphicx} 
\usepackage{dcolumn}  
\usepackage{bm}       
\usepackage{xcolor}

\begin{document}

\preprint{FM ZnO}

\title{Surface magnetization in non-doped ZnO nanostructures}

\author{A. L. Schoenhalz, J. T. Arantes, A. Fazzio and G. M. Dalpian}
\affiliation{
 $^1$Centro de Ci\^encias Naturais e Humanas, 
Universidade Federal do ABC, Santo Andr\'e, SP, Brazil\\
}
\date{\today}

\begin{abstract}
We have investigated the magnetic properties of non-doped ZnO nanostructures by using {\it ab initio} total energy
calculations. Contrary to many proposals that ferromagnetism in non-doped semiconductors
should be induced by intrinsic point defects, we show that ferromagnetism in nanostructured materials
 should be mediated by extended defects such as surfaces and grain boundaries. 
This kind of defects create delocalized, spin polarized states that should be able to warrant
long-range magnetic interactions. 
\end{abstract}

\keywords{nanocrystals, ZnO, surface}
 
\maketitle


The observation of high-temperature ferromagnetism in transition metal doped ZnO 
has triggered extensive studies on this kind of material.
However, the fundamental nature of the magnetic interactions between 
transition metal impurities, and the origin of the 
magnetism in ZnO is still under debate.
\cite{chambers_review,coey_review,pan_review}
Recent experimental results 
reported room-temperature ferromagnetism even without  the
inclusion of magnetic impurities, 
the sometimes called {\it phanton ferromagnetism}\cite{coey_jpd134012},
specially in wide 
bandgap oxides. This kind of phenomena was first reported by the group of Coey for HfO$_2$,\cite{coey_nature}
and followed by others\cite{lucio,hong_prb73p132404}.
Theoretical calculations proposed that ferromagnetism in this material
was due to the exhistence of intrinsic point defects.\cite{sanvito_prl} This proposal
has been put in check by recent calculations, since the population of these
defects should not be large enough to warrant a
magnetic coupling
between impurities.\cite{jorge_prb}.

Interestingly, ferromagnetism in non-doped materials is mostly observed in nanostructured samples composed of
nanoparticles and thin films
\cite{garcia_nl7p1489,sundaresan_prb74p161306,mendez_nl8p1562,hong_prb73p132404}.
There are several tentatives to try to explain these results, but they are
very controversial. Some works propose the ferromagnetic response
is due to oxygen vacancies \cite{sundaresan_prb74p161306,banerjee_apl91p182501}, while 
others say that it is due to zinc vacancies\cite{wang_prb77p205411,hong_jp07}. 
There are also results proposing that {\it intrinsic defects}\cite{xu_apl08}
or {\it interstitial zinc at the surface}\cite{yan_jpl92p081911} are
the responsible for the observed magnetic phases. As pointed out in Ref. \cite{jorge_prb}, 
the problem with these approaches is that the population of intrinsic defects 
is usually not large enough to reach the percolation limit,
and lead to a macroscopic magnetization, since the exchange interaction
between these impurities is expected to be short-ranged. Neither {\it double-exchange} nor
{\it superexchange}\cite{dalpian_bcm} are able to explain the observed
Curie temperatures in these materials.

It is clear, from the wide variety of experimental results, that magnetism 
in non-doped samples should be related to some kind of defect in the material. As point defects are 
very unlikely to be responsible for this magnetism, we turn our attention
to extended defects. Gamelin and co-workers \cite{daniel_jacs,daniel_jap} have shown 
that it is possible to tune the magnetic properties of transition-metal doped
TiO$_2$ by carefully controling the morphology of the grain boundaries in their 
samples. Garcia {\it et al} also showed that non-doped ZnO nanoparticles could have
a small ferromagnetic response, depending on the organic capping of their nanocrystals
\cite{garcia_nl7p1489}. Another recent work published by the Coey group\cite{coey_jpd134012}
suggests a charge-transfer mechanism
to explain ferromagnetism in oxide nanoparticles. In this model, 
electrons are transfered from the core of the nanocrystal to its surface, 
leading to a magnetization.  
In view of the large variety of controversial results on this subject, 
we have investigated, through {\it ab initio} calculations,
 the magnetic properties of surfaces in ZnO nanocrystals.
We have used ZnO nanocrystals with different
structures, geometries and shapes, and observed a 
magnetization at the surface of the nanocrystals that
should be responsible for ferromagnetism in these
materials.

Nanocrystals used in this study consist of an approximately spherical part cut from the 
bulk crystal.
We have analyzed nanocrystals
in the  wurtzite (WZ) and zinc blende (ZB) structures.
In order to generate the nanocrystals,\cite{dalpian_nl} we have to define a center, that can be located
either on an atom (AC) or a bond(BC). Bond-centered nanocrystals are stoichiometric, while
atom-centered nanocrystals not. We then define a radius,
and remove all atoms beyond the sphere defined by the center and the radius.
This kind of nanocrystals are usually  
passivated with fictitious hydrogen atoms, in order to remove dangling bonds from its surfaces
\cite{dalpian_nl}. In this work, we will not use this saturation, since our objective
is to study the effect of these surfaces. We have studied different 
sizes of nanocrystals for each structure (WZ and ZB). The smaller has a diameter
of $\sim0.9$nm and the larger one has a diameter of $\sim1.5$nm.
The surface of each nanocrystal reconstructed in a different way, giving us a 
broad range of results to model several different 
defects, including planar surfaces, steps, kinks, ad-islands, ad-atoms,
dimers and others. We believe that in our calculations we have taken into account 
the majority of possible motifs present at any nanocrystal surface.

Our simulations were performed using 
density-functional theory (DFT),\cite{klaus} employing the projected augmented wave method 
(PAW) as implemented in the VASP code\cite{vasp}.
Electronic exchange-correlation was treated using the local density approximation (LDA).
We have also tested the generalized gradient approximation (GGA) but we have not 
observed any changes to our conclusions within this approach.
We used a plane wave basis set, and the nanocrystals are separated by its images
by a vacuum region of 0.6 nm in all directions. All atoms were allowed to relax 
until the forces are smaller than 0.025 eV/\AA.

\begin{figure}
\includegraphics[scale=0.65]{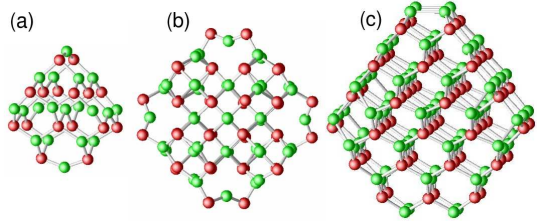}
  \caption{(Color online) Surface reconstruction of nanocrystals with (a) 35, (b) 87 and (c) 147 atoms
that present a spontaneous magnetization.
 }
\label{reconstruction}
\end{figure}

\begin{figure}
\includegraphics[scale=0.65]{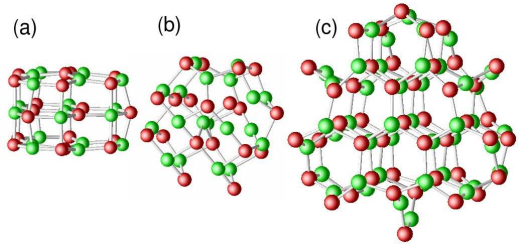}
  \caption{ (Color online) Surface reconstruction of nanocrystal with (a) 39, (b) 38 and (c) 88 atoms and where no spontaneous magnetization was observed.
 }
\label{reconstruction2}
\end{figure}

When we minimize the forces on all atoms of our bulk-like nanocrystals,
we observe that 
the surface atoms  
reorganize, through different reconstructions in the 
surface. Some of these reconstructions are shown in Fig. \ref{reconstruction} 
and \ref{reconstruction2} and include changes of the distance between the atoms 
in the surface, changes of the binding angle, and a 
strong tendency of the surface atoms to  form a ``graphitic'' reconstruction.
\cite{zhang_apl90p023115}. 
This ``graphitic'' structure has been reported in previous studies of ZnO
thin films and nanostructures.\cite{kulkarni_prl06,zhang_apl90p023115}. 

\begin{figure}
\includegraphics[scale=0.45]{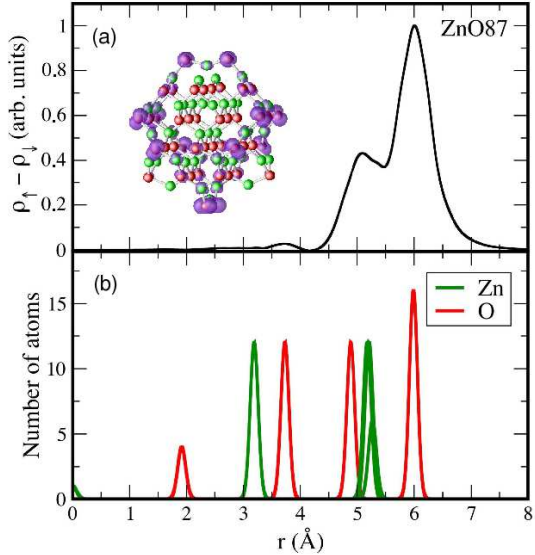}
  \caption{\label{structure}
(Color online) Magnetization for the 87-atoms nanocrystal. (a) Difference of $\rho_{\uparrow}-\rho_{\downarrow}$, showing that the magnetization is localized at the nanocrystal surface. (b) Radial distribution function, showing that the magnetization for this nanocrystal is mainly due the oxygen atoms of surface.
 }
\label{fig.dens.fdr}
\end{figure}

Although no magnetization is observed on the hydrogen-saturated nanocrystals, 
some of the non saturated nanocrystals show spontaneous magnetization without
magnetic impurities. This magnetization can be orginated from a
wide variety of different
reconstructions and motifs, such as zinc 
dimers, broken bonds, zinc and oxygen atoms with dangling bonds. 
Independent of  the reconstruction, the magnetization is always strongly localized at the
 nanocrystal 
surface. This can be observed in Fig. \ref{fig.dens.fdr}.
 In Fig. \ref{fig.dens.fdr}a, we
show the radial distribution function of the spin charge density, providing a quantitative comparison between the spin charge density at the center and at the surface of the nanocrystal.
For this specific nanocrystal, we observe that the magnetization is mainly localized at a radius of 6\AA. Fig. \ref{fig.dens.fdr}b
shows the radial distribution function of all atoms, 
confirming that the magnetization is localized in the outer atoms
of the nanocrystal(in this case, Oxygen atoms). 
In  the inset, we plot the spin charge density, i.e.,
the difference between spin-up and spin-down charge densities, for the 87-atom nanocrystal. We can observe
that the magnetization is more localized at the surface.

After analyzing all studied nanocrystals, we were not able to point  a single kind of defect 
as responsible for the magnetization.
We can not correlate the observation of the magnetization to 
a single defect like an O or a Zn atom at the surface, since the magnetization
can be observed in both cases. As extended defects might contain several different kinds
of point defects, we propose that extended defects
such as surfaces or grain boundaries should be the responsible to sustain 
macroscopic magnetizations in nanostructured samples. As extended defects can 
cover wide areas or volumes, they can mediate the long range 
interactions necessary for ferromagnetism to be observed in these samples.

Although we observed many motifs that present magnetization, the surface magnetization was not observed in all of our nanocrystals.
Interestingly, some kind of order inside the nanocrystal
was necessary for the magnetization to be observed: 
for nanocrystals whose overall structure was strongly reorganized, almost resembling an amorphous
structure, no magnetization was observed. 
In Fig. \ref{reconstruction2} we show some nanocrystals in which the 
spin-polarization is not observed. The change of the structure in the
smaller nanocrystals is clear from this figure. In Table I, we show a summary of the 
calculated magnetization for some nanocrystals.

\begin{table}[ht]
\centering
\caption{\label{tab02} 
Total magnetization for nanocrystals with different sizes and structures. ZB, WZ, BC and AC stand respectively
for Zinc Blende, Wurtzite, bond centered and atom centered. }
\begin{ruledtabular}
\begin{tabular}{ccc}
Number of atoms & Structure & Total magnetization \\  \hline\hline
35 & ZB-AC & 2 $\mu_B$ \\ 
87 & ZB-AC & 4 $\mu_B$ \\ 
147 & ZB-AC & 2 $\mu_B$ \\ 
38 & ZB-BC & 0 \\ 
86 & ZB-BC & 2 $\mu_B$ \\ 
238 & ZB-BC & 0 \\ 
39 & WZ-AC & 0 \\ 
92 & WZ-AC & 2 $\mu_B$ \\ 
34 & WZ-BC & 0 \\ 
88 & WZ-BC & 0
\end{tabular}
\end{ruledtabular}
\label{table2}
\end{table}

 These results are in agreement with recent 
experimental results, confirming that ZnO nanostructures can be magnetic without 
transition metals.
Of particular interest is the  work of Garcia {\it et al} \cite{garcia_nl7p1489}
that shows that, depending on the capping molecules, different strengths of 
magnetization can be observed. In our calculations the magnetic moment per surface atom
can vary from 0.01$\mu_b$ to 0.04$\mu_b$, depending on the nanocrystal size.
Larger nanocrystals have smaller magnetizations per surface atom. 
These results are in reasonable agreement with the experimental results \cite{garcia_nl7p1489}
that report a magnetization of the order of 0.001$\mu_b$ for larger nanocrystals.

Following the proposal of Coey {\it et al}\cite{coey_jpd134012} of a charge transfer 
magnetism, we have also analyzed
the charge transfer from the center to the surface of our nanocrystals. 
Our results indicate that, if present, the charge transfer will be extremelly
small. In some cases, the charge at the surface atoms is even smaller than the total
charge at the central ones, showing that there should not be any charge transfer in 
neutral nanocrystals.
We propose that depending on the surface reconstruction, surface states might be created
in the energy gap of the nanocrystal.
These surface states may be exchange splitted, leading to a net macroscopic magnetization
since these states are delocalized through the whole nanocrystal. The magnetization of these 
surface states will be responsible for the ferromagnetic response in nanostructured materials.

In conclusion, we have analyzed the magnetization of small non-saturated ZnO 
nanocrystals by looking for several different surface  reconstructions.
We propose that point deffects are not responsible for this magnetization. 
We show that, depending on the structure, the whole nanocrystal surface might
create delocalized levels that are spin-polarized
and leading to a macroscopic, long-range magnetization in the samples. 
Ferromagnetism was not observed in all of our nanocrystals, what is also in
agreement with experiment: not all experimental results report ferromagnetism in non-doped samples,
even for samples produced with the same procedure. 
Following in this direction, other extended defects that insert delocalized levels
in the gap of the material also might be important to understand the magnetization in non-doped
samples. These include dislocations, grain boundaries and interfaces. 
Our findings also should be important to explain the magnetization in other 
non-doped materials such as HfO$_2$ and TiO$_2$.

This work was supported in part by brazilian agencies
CAPES, FAPESP and CNPq.


\end{document}